\documentclass[twocolumn,superscriptaddress,pre]{revtex4-1}

\newcommand{\Tc}{\ensuremath{T_\mathrm{c}}\xspace}
\newcommand{\NR}{\ensuremath{N_\mathrm{R}}\xspace}
\newcommand{\NS}{\ensuremath{N_\mathrm{S}}\xspace}
\newcommand{\sigmaS}{\ensuremath{\sigma_\mathrm{S}}\xspace}
\newcommand{\sigmaR}{\ensuremath{\sigma_\mathrm{R}}\xspace}

\usepackage[T1]{fontenc}
\usepackage[utf8]{inputenc}
\usepackage{graphicx}
\usepackage{amsmath}
\usepackage{amssymb}
\usepackage{bm}
\usepackage{color}
\usepackage{xspace}
\usepackage{hyperref}
\usepackage{soul}

\renewcommand{\vec}[1]{\bm #1}
\makeatletter
\newcommand*{\balancecolsandclearpage}{%
  \close@column@grid
  \cleardoublepage
  \twocolumngrid
}
\makeatother
\graphicspath{{./figures/}{./}}

\begin{document}

\title{Dimensional crossover in the aging dynamics of spin glasses in a film geometry}
\author{L.A.~Fernandez}\affiliation{Departamento  de F\'\i{}sica Te\'orica,
  Universidad Complutense, 28040 Madrid, Spain}\affiliation{Instituto de
  Biocomputaci\'on y F\'{\i}sica de Sistemas Complejos (BIFI), 50018 Zaragoza,
  Spain}

\author{E.~Marinari}\affiliation{Dipartimento di Fisica, Sapienza
  Universit\`a di Roma, INFN, Sezione di Roma 1, and CNR-Nanotec,
  I-00185 Rome, Italy}

\author{V.~Martin-Mayor}\affiliation{Departamento  de F\'\i{}sica Te\'orica, Universidad Complutense, 28040 Madrid, Spain}\affiliation{Instituto de Biocomputaci\'on y F\'{\i}sica de Sistemas Complejos (BIFI), 50018 Zaragoza, Spain}

\author{I.~Paga}\affiliation{Dipartimento di Fisica, Sapienza
  Universit\`a di Roma, INFN, Sezione di Roma 1,Italy}\affiliation{Departamento  de F\'\i{}sica Te\'orica,
  Universidad Complutense, 28040 Madrid, Spain}

\author{J.J.~Ruiz-Lorenzo}\affiliation{Departamento de F\'{\i}sica, Universidad de Extremadura, 06006 Badajoz, Spain}\affiliation{Instituto de Computaci\'on Cient\'{\i}fica Avanzada (ICCAEx), Universidad de Extremadura, 06006 Badajoz, Spain}\affiliation{Instituto de Biocomputaci\'on y F\'{\i}sica de Sistemas Complejos (BIFI), 50018 Zaragoza, Spain}

\date{\today}

\begin{abstract}
Motivated by recent experiments of exceptional accuracy, we study
numerically the spin-glass dynamics in a film geometry. We cover all
the relevant time regimes, from picoseconds to equilibrium, at
temperatures at and below the 3D critical point.  The dimensional
crossover from 3D to 2D dynamics, that starts when the correlation
length becomes comparable to the film thickness, consists of four
dynamical regimes. Our analysis, based on a Renormalization
Group transformation, finds consistent the overall physical picture employed
by Orbach et al. in the interpretation of their experiments.
\end{abstract}

\maketitle

\section{Introduction} Spin glass physics \cite{mezard:87,fisher:91} has
interested, puzzled and motivated the scientific community in the last
fifty years, and it is still full of open challenges. The models
behind this approach are both of dramatic theoretical and
computational interest and of widespread potential interest, since
they describe very different systems and situations.  Glassy physics
and the outstanding problem of the explanation of the amorphous state
can receive important clarifications from the ideas developed in this
context. Besides, very diverse fields like neuroscience, optimization,
active matter, protein folding or DNA and RNA physics are turning out
to be connected to the field and, indeed, progress thanks to the same
techniques~\cite{young:98}.

In the lab, spin-glass samples are permanently out of equilibrium when
studied at temperatures below the critical one, $\Tc$, implying that
the equilibrium theory is not always sufficient. A possible approach
to overcome this difficulty is extracting from the non-equilibrium
dynamics crucial information about the (so difficult to reach)
equilibrium
regime~\cite{cugliandolo:93,franz:99,janus:10b,janus:16}. However,
custom-built computers~\cite{janus:14} and other simulation
advances~\cite{manssen:15,fernandez:15} have made it possible to study
theoretically~\cite{janus:08b,janus:09b,manssen:15b,janus:17b,janus:18,fernandez:19,fernandez:18b}
the simplest experimental protocol. In this protocol, see
e.g.~\cite{joh:99}, a spin glass at some very high-temperature is
fastly quenched to the working temperature $T<\Tc$ and the
excruciatingly slow growth of the spin-glass correlation length $\xi$
is afterwards studied as a function of the time elapsed since the
quench, $t$. Although simulations do not approach yet the experimental
time and length scales ($t\sim 1$ hour and $\xi\sim 100\, a_0$, where
$a_0$ is the average distance between magnetic moments), the range
covered is already significant: from picoseconds to
milliseconds~\cite{manssen:15,fernandez:15} or even 0.1 seconds using
dedicated computers~\cite{janus:08b,janus:18} (or conventional ones in
the case of two-dimensional spin
glasses~\cite{fernandez:19,fernandez:18b}).

Yet, thanks to advances in sample preparation, a new and promising
experimental protocol have appeared in the last five years. Indeed,
single-crystal spin glass samples with a thin-film geometry (thickness
of $4.5 - 20$ nm) have been
investigated~\cite{guchhait:14,guchhait:17,zhai:17,kenning:18}.  These
experiments are interpreted in terms of a correlation length $\xi$
saturating at a constant value after reaching a characteristic length
scale, namely the thickness of the film. The bounded growth of $\xi$
along the longitudinal direction of the film is a direct experimental
confirmation~\cite{guchhait:14} for a lower critical dimension $2<
D_{\text{l}}^{\text{c}} <3$, in agreement with the theoretical
expectation $D_{\text{l}}^{\text{c}} \sim
2.5$~\cite{bray:86,franz:94,maiorano:18}. The film-geometry has
allowed as well for extremely accurate measurements~\cite{zhai:17} of
the aging rate
\begin{equation}\label{eq:aging-rate-def}
  z(T,\xi)=\frac{\mathrm{d}\,\log t}{\mathrm{d}\,\log\xi}\,,
\end{equation}
which gives access to the dominant free-energy barrier $\varDelta$,
$t\sim \tau_0\,\text{exp}[\varDelta/(k_{\text{B}} T)]$~\footnote{At
  the critical temperature $\Tc$, the aging rate coincides with the
  so-called dynamic critical exponent}
[$\tau_0=\hbar/(k_{\text{B}}\Tc)$ is a time scale]. The increased
accuracy has shown that, contrary to previous
expectations~\cite{joh:99,janus:08b,janus:09b,janus:17b}, the aging
rate depends on $\xi$ (see also~\cite{bert:04,janus:18}). Besides, the
dependency of the barrier $\varDelta$ with the applied magnetic field
has been clarified~\cite{guchhait:17}. However, a theoretical study of
these fascinating thin-film experiments is lacking.

Here, we investigate the spin glass dynamics in a film geometry through
large-scale numerical simulations. We analyze the dimensional-crossover and we
critically assess the hypothesis of a dynamical arrest that becomes complete
as soon as transversal saturation of the correlation length happens.  Somewhat
surprisingly, we find a rich dynamic behavior with no less than four different
regimes (3D growth at short times, a double crossover regime with a faster
growth for intermediate times, and a final equilibration regime). We analyze
our results by combining the phenomenological Renormalization
Group~\cite{nightingale:76} with recent analysis of the two-dimensional spin
glass dynamics~\cite{fernandez:19,fernandez:18b}. On the light of our
results, the interpretation of thin-film
experiments~\cite{guchhait:14,guchhait:17,zhai:17,kenning:18} seems
essentially correct, albeit slightly oversimplified.

The remaining part of this work is organized as follows. In Sect.~\ref{sect:2-3D-physics} we recall the spin glass physics in $2\text{D}$ and $3\text{D}$.
In Sect.~\ref{sec:model} we define the model and provide details about our simulation and our analysis protocol.
Our main results are given in Sect.~\ref{sec:results}, where we discovered a dynamics characterized by four aging regimes and through the Renormalization Group approach we found a non-trivial temperature mapping between a film and a $\text{2D}$ system.
Finally, we provide our conclusions in Sect.~\ref{sec:end}. Further details are provided in the appendices.

\section{2D and 3D spin glass dynamics}
\label{sect:2-3D-physics}
Before addressing the dimensional crossover, let us recall a few crucial facts about the
very different dynamic behavior of spin glasses in spatial dimensions
$D=2$~\cite{fernandez:19,fernandez:18b} and $D=3$~\cite{janus:18}.

In 3D, a phase transition at $T\!=\!\Tc$ separates the high-temperature
paramagnetic phase from the spin-glass
phase~\cite{gunnarsson:91,palassini:99,ballesteros:00}. The aging
rate~\eqref{eq:aging-rate-def} is $\xi$-independent at exactly
$T=\Tc$, which results into a power-law dynamics $\xi\sim
t^{1/z(\Tc)}$, with $z(\Tc)=6.69(6)$.  At $T<\Tc$, but only once $\xi$
grows large-enough~\cite{janus:18}, the aging-rate grows with
$\xi$ (the dynamics slows-down, and a
power-law description is no longer appropriate). A simplifying feature
is that the renormalized aging-rage $z(T,\xi)T/\Tc$ is roughly
$T$-independent: when $T<\Tc$, the dominant barrier $\varDelta(\xi)$
depends little (or not at all) on temperature.

In 2D, we are in the paramagnetic phase for any $T>0$. Hence,
$\xi(t,T)$ eventually reaches its equilibrium limit
$\xi_{\text{eq}}(T)$, which can be \emph{very}
large~\cite{fernandez:16,fernandez:19}: for $T\to 0$,
$\xi_{\text{eq}}(T)\propto 1/T^{\nu_{\text{2D}}}$,
$\nu_{\text{2D}}=3.580(4)$~\cite{khoshbakht:17}. When
$a_0\ll\xi(t,T)\ll\xi_{\text{eq}}(T)$ we have a power law $\xi\propto
t^{1/z_{\text{2D}}}$, with $z_{\text{2D}}\approx 7.14\,$ irrespective of
$T$~\cite{fernandez:19}: 2D dynamics may be much faster than 3D
dynamics (aging rates $z\sim 15$ are not uncommon in 3D at low
$T$). For times scale $t\gg \tau^{\text{2D}}_{\text{eq}}(T)$ equilibrium is
approached. A super-Arrhenius behavior is found for
$\tau^{\text{2D}}_{\text{eq}}(T)\propto
\text{exp}[\varDelta^{\text{2D}}(\xi_{\text{eq}})/T]$, where the barrier
$\varDelta^{\text{2D}}(\xi_{\text{eq}})$ grows very mildly with
$\xi_{\text{eq}}$~\cite{fernandez:19}.

\section{Model and protocol}
\label{sec:model}
We consider the Edwards-Anderson model~\cite{edwards:75} in a cubic
lattice with a film geometry.  Our films have two long sides of length
$L_x=L_y$, and thickness $L_z\ll L_x$ (in the experiments, $L_z$
ranges from 8 to 38 layers~\cite{zhai:17}). We impose periodic
boundary conditions (PBC) along the two longitudinal directions $X$
and $Y$. We have simulated $L_x=256$ and $L_z=4$, 6, 8 and 16.  We
always keep $L_x\gg\xi$, in order to effectively take the
$L_x\to\infty$ limit. On the other hand, we have considered both PBC
and open boundary conditions (OBC) along the short transversal
direction $Z$. For simplicity, we discuss here only PBC
[see appendix C for the qualitatively similar OBC results].

At the initial time $t=0$ our fully disordered films are abruptly
quenched down to the working temperature $T$, which we simulate with
Metropolis dynamics ($t$ is measured in full-lattice sweeps, a sweep
roughly corresponds to 1 picosecond~\cite{mydosh:93}).  Our
$\sigma_{\vec x}=\pm 1$ spins interact with their lattice
nearest-neighbors through a Hamiltonian $H_\mathrm{EA} = - \sum_{<\vec
  x,\vec y>} J_{\vec x\vec y}\, \sigma_{\vec x}\, \sigma_{\vec y}$,
where the quenched disordered couplings are $\{ J_{\vec x\vec y} \} =
\pm 1$ with $50\%$ probability.  For each quenched realization of the
coupling (a sample) we study $N_\mathrm{R}$ real replicas.
$N_\mathrm{R}$ has been selected for optimal performances
(see appendices A and B for further details).

The spatial autocorrelation function \cite{janus:09b} is defined as
$C_4\left(T,\vec r,t\right) = \overline{\left\langle
  q^{(a,b)}\left(\vec x, t\right) q^{(a,b)}\left(\vec x+\vec r,
  t\right)\right\rangle_T }\,,$ $q^{(a,b)}\left(\vec x, t\right)
\equiv \sigma^{(a)} \left( \vec x,t\right) \sigma^{(b)} \left(\vec
x,t\right)\,,$ where the indices $(a,b)$ label the different real
replicas, $\overline{(\cdots )}$ denotes the average over the disorder
and $\langle \cdots\rangle_T $ stands for the average over the thermal
noise at temperature $T$.

For the \emph{longitudinal} lattice displacements $\vec r=(r,0,0)$ or
$(0,r,0)$, one expects \cite{parisi:88,fernandez:18b}
\begin{equation}
  C_4\left(T,\vec r,t\right) \sim
  \frac{f(u,v)}{r^{\theta}}\,,\ u=\frac{r}{\xi^{\parallel}(T,t)}\,,\ 
  v=\frac{\xi^{\parallel}(T,t)}{\xi^{\parallel}_{\text{eq}}(T)} \; ,
  \label{fC4}
\end{equation}
where $f(u,v)$ is an unknown scaling function~\footnote{A
  Renormalization Group argument implies that the
  scaling function $f(u,v)$ depends as well on the effective
  \emph{two-dimensional} temperature $T_{\text{eff,2D}}$, see
  Eq.~\eqref{PRG-eq}. In equilibrium, $f(u,v\!=\!1)$ decays for large
  $u$ as $\exp(-u)/\sqrt{u}$~\cite{fernandez:18b} ($v=1$ is reachable
  in a film at $T<\Tc$ only thanks to the 3D-to-2D
  crossover~\cite{guchhait:14}). Off-equilibrium, $f(u,v\!<\!1)$
  decays super-exponentially in $u$~\cite{fernandez:18b}.}.
Fortunately, we can study the dynamical growth of $\xi^{\parallel}$
without parameterizing $f(u,v)$ through the integral estimators
\cite{janus:08b,janus:09b} $I_k(T,t) = \int_0^{\infty} \mathrm{d}r \, r^k
C_4(T,r,t)$: $\xi^{\parallel}_{k,k+1}(T,t)\equiv
I_{k+1}(T,t)/I_k(T,t)$.  We shall specialize to
$\xi^{\parallel}_{12}(T,t)$ which has been thoroughly
studied~\cite{janus:18,fernandez:19,fernandez:18b}.

As for correlations along the short transverse direction, we obtain
another  characteristic length $\xi^{\perp}$ through:
\begin{equation}
\xi_{12}^{\perp} = \sum_{r=0}^{L_z/2} r^2\,C_4^{\perp}(T,r,t)\,\Big/\, \sum_{r=0}^{L_z/2} r\,C_4^{\perp}(T,r,t)\,,
\label{sum_estimator}
\end{equation}
(the sum is truncated at half of the transversal thickness because
of the PBC).  Also in this
case we use $k=1$.  We show $\xi_{12}^{\parallel}$
in Fig.~\ref{fig:aging_rate} and $ \xi_{12}^{\perp}$ in
Fig.~\ref{fig:longitudinal_vs_transverse}, for $T=1.1\approx\Tc$, $T=0.98\approx 0.89\Tc$ and $T=0.7 \approx 0.64\Tc$.

\section{Results}
\label{sec:results}
\begin{figure}
\centering
\includegraphics[width=\columnwidth]{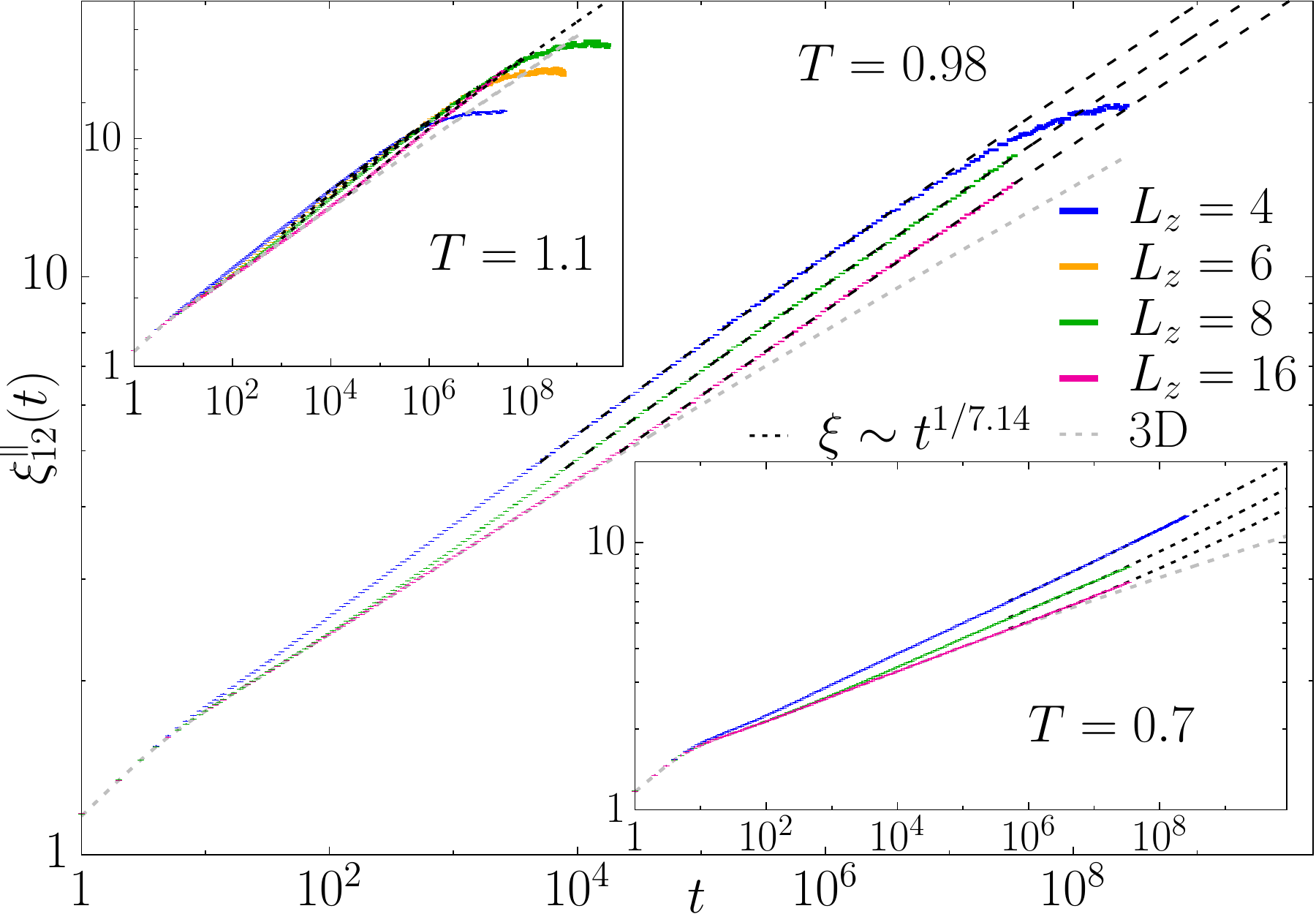}
\caption{\label{fig:aging_rate}The longitudinal correlation length
  $\xi_{12}^{\parallel}(T,t)$, as computed in films of thickness $L_z$,
  versus the waiting time $t$ after a quench to temperature $T$, for
  $T=0.98$ ({\bf main}), $T=1.1$ ({\bf upper inset}) and $T=0.7$ ({\bf lower
    inset}). The critical temperature is $\Tc=1.102(3)$~\cite{janus:13}.  As a
  reference, we also show purely 3D dynamics (data taken from
  Ref.~\cite{janus:18}) and fits to 2D dynamics
  $\xi_{12}^{\parallel}(L_z,T,t)\approx b(L_z,T) \ +\ a(L_z,T)
  t^{1/z_\mathrm{2D}}$, with $z_\mathrm{2D}=7.14\,$~\cite{fernandez:19} 
  [fit parameters:  $b(L_z,T)$ and $a(L_z,T)$].}
\end{figure}

Let us start by considering the longitudinal $\xi^{\parallel}$ in
Fig.~\ref{fig:aging_rate}. All the main points can be assessed by
looking at the data at $T=0.98$: the data at $T=0.7$ and $T=1.1$ are
useful to confirm this picture where four different regimes of
interest appear.  In the first regime, for small times, the growth of
the $\xi_{12}^{\parallel}$ is indistinguishable from what happens in
3D. Eventually the growth rate changes (for example for $T=0.98$ and
$L_z=16$ at a time larger than $10^4$) and the system enters a second
regime where $\xi^{\parallel}$ grows faster than in $D=3$. After this
transient for a while, in a third regime, $\xi^{\parallel}$ grows like
in 2D which, as we explained above, for $T<\Tc$ is a faster-than-3D
growth. Finally, the fourth regime corresponds to the saturation of
$\xi^{\parallel}$ to its equilibrium value (the 4th regime is
completed in our data for $L_z=4$ at $T=0.98$, and for all our $L_z$ at
$T=1.1$).

\begin{figure}
\centering
\includegraphics[width=\columnwidth]{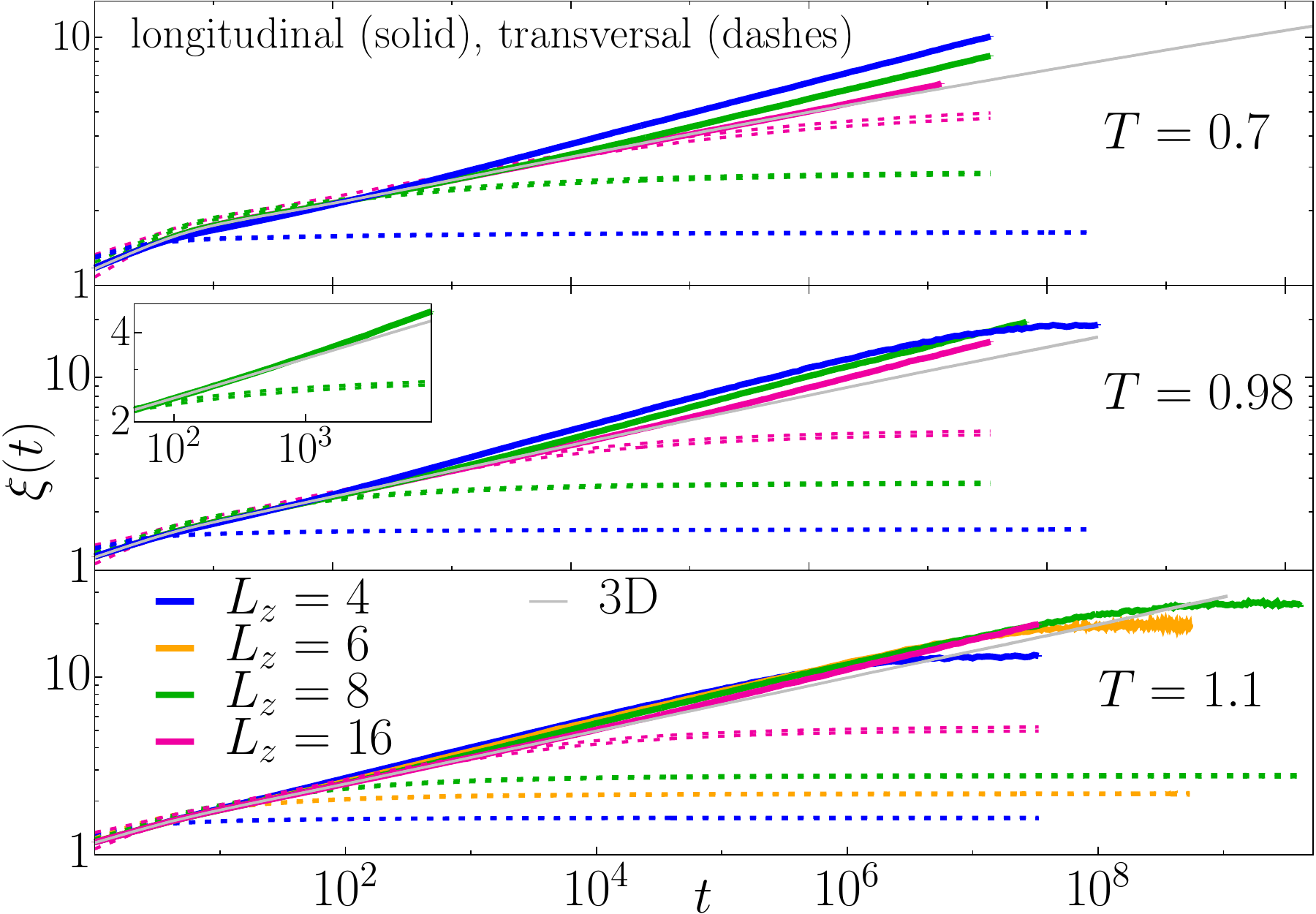}
\caption{\label{fig:longitudinal_vs_transverse} Growth of the
  longitudinal $\xi_{12}^{\parallel}$ (solid lines) and of the
  transversal $\xi_{12}^{\perp}$ (dashes lines) correlation lengths
  with the waiting time $t$ after a quench to temperature $T$. The
  inset (for $L_z=8$) is a zoom of the saturation of
  $\xi_{12}^{\perp}$ and of the separation between the
  $\xi_{12}^{\parallel}$ and the bulk correlation length (see the main
  text for more details).}
\end{figure}

Next, we compare $\xi^{\parallel}$ and $\xi^{\perp}$ in
Fig.~\ref{fig:longitudinal_vs_transverse}.  The dynamical behaviors of
these two quantities are very different.  As expected $\xi^{\perp}$
saturates to a value near $L_z/2$ (which is the maximum value with
PBC). However, $\xi^{\parallel}$ continues growing \emph{after}
$\xi^{\perp}$ saturates: in no way the time where $\xi^{\perp}$ and
$\xi^{\parallel}$ stop growing is the same.  In fact,
$\xi_{12}^{\parallel}$ needs time to respond to the saturation of
$\xi_{12}^{\perp}$: even the switch from the 3D like growth to the
faster-than-3D growth arrives at a later time (see the inset in the
$T=0.98$ part of Fig.~\ref{fig:longitudinal_vs_transverse}).
Saturation of $\xi^{\parallel}$ eventually happens, at later
times. Although $\xi_{12}^{\parallel}$ saturates as well, these two
time scales are remarkably different.

\begin{figure}
\centering
\includegraphics[width=\columnwidth]{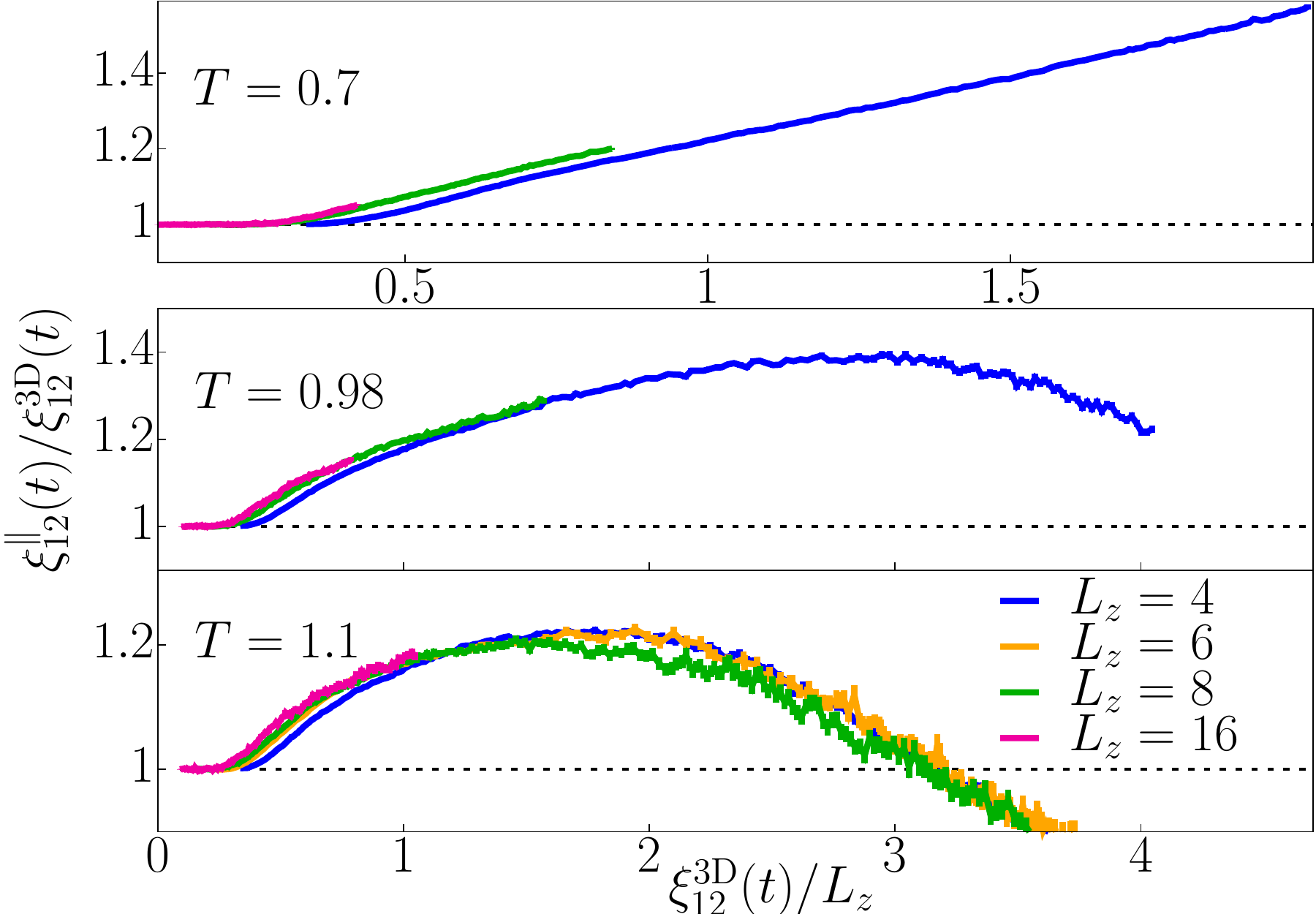}
\caption{\label{fig:scaling_function_xi12} Dynamical scale-invariance
  for the dimensionless quantity $ \xi_{12}^\mathrm{film}(t)/
  \xi_{12}^\mathrm{3D}(t)$ as a function of the rescaled bulk length
  $\xi_{12}^\mathrm{3D}(t)/ L_z $.}
\end{figure}

In order to gain some understanding, we have identified a second
characteristic length (besides the thickness $L_z$) that controls the
3D-to-2D crossover, namely the bulk correlation $\xi_{12}^\mathrm{3D}(t)$~\cite{janus:18}. We
have studied the behavior of the dimensionless
$\xi_{12}^\parallel(t)/\xi_{12}^\mathrm{3D}(t)$ as a function of
$\xi_{12}^\mathrm{3D}(t)/L_z$. In other words, we change variables from $t$ to
$\xi_{12}^\mathrm{3D}(t)$.  As one can see in Fig.~\ref{fig:scaling_function_xi12} a
very good scaling behavior emerges. This not only confirms the existence of the
3D-to-2D crossover, but also unveils some of its features.  Indeed, the ratio
$\xi_{12}^\parallel(t)/ \xi_{12}^\mathrm{3D}(t)$ grows beyond  1, thus
signaling a faster-than-3D dynamics as soon as $\xi_{12}^\mathrm{3D}(t)\approx
L_z/4$ (for all our $T<\Tc$).

The scale-invariance evinced in Fig.~\ref{fig:scaling_function_xi12}
prompts us to consider the film dynamics from the
Renormalization-Group perspective (see e.g.~\cite{amit:05}). Indeed,
in equilibrium, phenomenological renormalization~\cite{nightingale:76}
maps our film at temperature $T$ to a truly 2D spin glass at an
effective temperature $T_{\text{eff,2D}}$ (for details, see below
and appendix D):
\begin{equation}\label{PRG-eq}
  \xi_{12}^{\parallel,\mathrm{eq}} \left(T, L_z\right) =
  L_z\, \xi_{12}^\mathrm{eq,2D} \left(T_{\text{eff,2D}}\right)\,,
\end{equation}
where the equilibrium correlation length $\xi_{12}^\mathrm{eq,2D}$ is
a smooth function of $T_{\text{eff,2D}}$ (provided that
$T_{\text{eff,2D}}>0$). For any fixed $T>\Tc$, $T_{\text{eff,2D}}$
increases with $L_z$ ($T_{\text{eff,2D}}\to \infty$ when
$L_z\to\infty)$. On the other hand, holding fixed $T\leq\Tc$ while $L_z$
grows, $T_{\text{eff,2D}}$ reaches a limit. The limit is neither $0$
nor $\infty$, because the whole spin-glass phase is critical in
3D~\cite{janus:10b,contucci:09}.

Two questions naturally appear: (i) \emph{Is the equilibrium
  mapping~\eqref{PRG-eq} meaningful for an aging, off-equilibrium
  film?\/} and (ii) \emph{Is it sensible to assume
  $T_{\text{eff,2D}}\approx T$?\/} (an assumption that, although not
explictly, underlies the experimental
analysis~\cite{guchhait:14,guchhait:17,zhai:17,kenning:18}).

\begin{figure}
\centering
\includegraphics[width=\columnwidth]{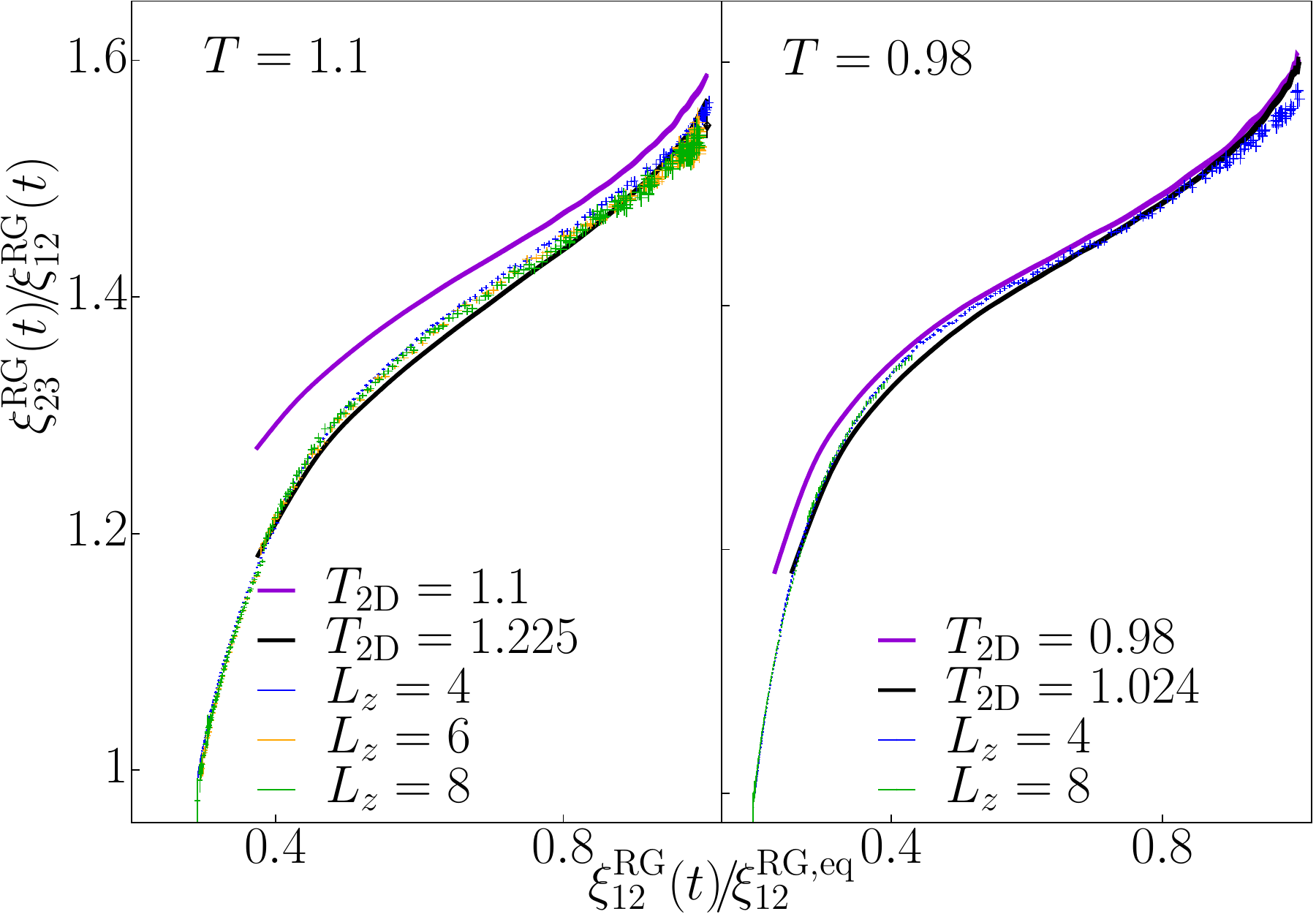}
\caption{\label{fig:scaling_function_xi23} The scale-invariant ratio
  $\xi^{\text{RG}}_{23}(T,t) / \xi^{\text{RG}}_{12}(T,t)$, versus the
  ratio $\xi^{\text{RG}}_{12}(t,T) / \xi_{12}^\mathrm{RG,eq}(T)$ as
  computed from the block-spins
  [$\xi^{\text{RG}}_{12}(t,T)$ grows monotonically to its equilibrium
  value $\xi_{12}^\mathrm{RG,eq}(T)$]. For $T=1.1\approx \Tc$ and
  $T=0.98\approx 0.9\Tc$, we compare the scaling function obtained
  from block-spins (as extracted from films of several thickness
  $L_z$), with two analogous functions computed in purely 2D systems.
  If the 2D system is considered at the film's temperature
  $T_{\text{2D}}=T$, the scaling function
  $\xi^{\text{2D}}_{23}(T,t)/\xi^{\text{2D}}_{12}(T,t)$ clearly
  differs from the block-spin result. On the other hand, the film and
  the 2D scaling function essentially coincide if the 2D system is
  considered at the effective temperature $T_{\text{eff,2D}}$ defined
  by Eq.~\eqref{PRG-eq}.}
\end{figure}

In order to address the above two questions, we perform on our aging
films a linear Kadanoff-Wilson block spin transformation of size $L_z$
(see appendix D): from $L_z^3$ of our original spins at time
$t$, we obtain a single renormalized spin in the renormalized 2D
system. The correlation functions computed for the \emph{aging}
renormalized spins can be compared with those of a truly 2D system at
the temperature $T_{\text{eff,2D}}$ obtained from
Eq.~\eqref{PRG-eq}. In particular, we have found it useful to compute
the dimensionless ratio
$\xi^{\text{RG}}_{23}(t)/\xi^{\text{RG}}_{12}(t)$ as computed from the
block spins, see Fig.~\ref{fig:scaling_function_xi23} (of course, for
the truly 2D system, $L_z=1$, $\xi$ and $\xi^{\text{RG}}$ are the same
quantity). This ratio is a smooth function of
$\xi^{\text{RG}}_{12}(t)/\xi_{12}^{\text{RG,eq}}$~\footnote{We use
  $\xi^{\text{RG}}_{12}(t)/\xi_{12}^{\text{RG,eq}}$ as a computable
  proxy for the unknown $v$ in Eq.~\eqref{fC4}, see
  Fig.~\ref{fig:scaling_function_xi23} and
  Ref.~\cite{fernandez:18b}.}\,].

As expected for a film at $T\leq \Tc$, the scaling function in
Fig.~\ref{fig:scaling_function_xi23} has no dependency in $L_z$. To be
precise, for $T=0.98$ we did not reach equilibrium in the $L_z=8$
film. However, by taking $\xi_{12}^{\text{eq}}$ from the block-spins
formed from the $L_z=4$ film, we find an excellent scaling:
corrections to scaling, if any, are not measurable within our
statistical accuracy for the films. 

Now, the very same scaling function can be computed in a truly 2D system at
temperature $T_{\text{2D}}$.  If one takes $T_{\text{2D}}=T$ ($T$ is the
film's temperature), we find a clear discrepancy in
Fig.~\ref{fig:scaling_function_xi23}~\footnote{The reader might be puzzled
  because all curves with $L_z>1$ in Fig.~\ref{fig:scaling_function_xi23} were
  obtained at the film's temperature $T$. Indeed, by taking the limit $L_z\to
  1$, one could (wrongly) conclude $T_{\text{eff,2D}}=T$. The way out of the
  paradox is, actually, one of the crucial ideas from the Renormalization
  Group~\cite{amit:05}: although the film gets mapped into a 2D spin glass,
  the corresponding 2D model is certainly not as simple as ours (square
  lattice, nearest-neighbors interaction, couplings $J=\pm
  1$). Phenomenological renormalization (PR)~\cite{nightingale:76} was
  invented, precisely, to keep using the simplified model at the prize of
  changing parameters such as temperature, hence the need for the
  $T_{\text{eff,2D}}$ from Eq.~\eqref{PRG-eq}. PR becomes exact only if
  $L_z\to\infty$ (rather than $L_z\to 1$).}. On the other hand, if we take
$T_{\text{2D}}=T_{\text{eff,2D}}$ the matching with the film's scaling
function is much better, in spite of the fact that corrections to scaling for
the 2D system are suppressed only when $T_{\text{2D}}\to
0$~\cite{fernandez:16b}.  Hence, the answer to our first question above is
\emph{yes, Eq.~\eqref{PRG-eq} is meaningful in the off-equilibrium regime, as
  well}.

As for our second question, Finite Size Scaling (see
e.g.~\cite{amit:05}) implies $\mathrm{d} T_{\text{eff,2D}}/\mathrm{d}
T\propto L_z^{1/\nu}$ at $\Tc$. Hence, when $L_z$ grows, the mapping
$T \to T_{\text{eff,2D}}$ becomes singular at $T=\Tc$. On the other
hand, we do not see questions of principle implying a singular mapping
for $T<\Tc$. Accordingly, we find $T_{\text{eff,2D}} \approx 1.11 T$
at $T=\Tc$, but $T_{\text{eff,2D}} \approx 1.04 T$ at $T\approx
0.9\Tc$.  In other words, \emph{the assumption $T_{\text{eff,2D}}
  \approx T$ is sensible}, provide that $T<\Tc$.

\section{Conclusions }
\label{sec:end}
Recent experiments in films~\cite{guchhait:14,guchhait:17,zhai:17,kenning:18}
focused on the saturation time, when the spin-glass correlation length $\xi$
no longer grows due to the dimensional crossover. From Eq.~\eqref{PRG-eq} and
Figs.~\ref{fig:scaling_function_xi23} and~\ref{fig:scaling_function_xi12}, we
expect for this saturation time
\begin{equation}\label{eq:discussion}
t_\text{sat}(L_z,T)\approx t_\text{3D}(L_z,T)
\,\varphi(\xi_{12}^\mathrm{eq,2D})\,
\tau^{\text{2D}}_{\text{eq}}(T_{\text{eff,2D}})\,,
\end{equation}
where $t_\text{3D}(L_z,T)$ is the time that a bulk, 3D system needs to reach
$\xi^\text{3D}_{12}=L_z$ at temperature $T$, $\varphi$ is a smooth function and
$\xi_{12}^\mathrm{eq,2D}$ is the correlation length of the effective 2D
system~\eqref{PRG-eq}. Hence, $t_\text{sat}$ is the product of the
renormalized time-unit $t_\text{3D}(L_z,T) \,\varphi(\xi_{12}^\mathrm{eq,2D})$,
times $\tau^{\text{2D}}_{\text{eq}}$ (i.e. the number of time-units that a
$2D$ system needs to equilibrate at the effective temperature
$T_{\text{eff,2D}}$). Experiments~\cite{guchhait:14,guchhait:17,zhai:17,kenning:18}
aim to extract the aging rate~\eqref{eq:aging-rate-def}, which depends on
$t_\text{3D}$ and $\xi_\text{3D}$, but they actually measure $t_\text{sat}$
and $L_z$. Nevertheless, we conclude that the experimental determination of the
aging rate is safe, thanks to three fortunate facts: (i) $T_{\text{eff,2D}}\approx T$
below $\Tc$, (ii) the factor $\varphi(\xi_{12}^\mathrm{eq,2D})$ depends only on
temperature (and very mildly so, see Fig.~\ref{fig:scaling_function_xi12}) and
(iii) the growth of $\tau^{\text{2D}}_{\text{eq}}$ is only \emph{slightly}
super-Arrhenius (the aging rate is blind to Arrhenius time-growth).

We remark as well that there is more than the saturation time in film dynamics
(we have identified four separate regimes). The exploration of this rich
behavior opens an opportunity window for the fruitful interaction of
experimental and numerical work in spin glasses.
 
\begin{acknowledgments}
We thank R. Orbach and G. Parisi for encouraging discussions.  This
work was partially supported by Spain's Ministerio de Econom\'ia,
Industria y Competitividad (MINECO) through Grants
No. FIS2015-65078-C2, No. FIS2016-76359-P (also partly funded by the
EU through the FEDER program), by Agencia Estatal de Investigaci\'on
(AEI) through Grant No. PGC2018-094684-B-C21 (also partly funded by
FEDER), by the Junta de Extremadura (Spain) through Grant No.
GRU10158 and IB16013 (both partially funded by FEDER), by the European
Research Council (ERC) under the European Union's Horizon 2020
research and innovation program (Grant No. 723955 - GlassUniversality
and Grant No. 694925 - LoTGlasSy) and by the Italian Ministery for
Education, University and Research (MIUR) through the \emph{FARE}
project \emph{Structural DIsorder and Out-of-Equilibrium Slow Dynamics
  in Interdisciplinary Applications}. Our simulations were carried out
at the BIFI supercomputing center (using the \emph{Cierzo} cluster)
and at ICCAEx supercomputer center in Badajoz (\emph{Grinfishpc} and
\emph{Iccaexhpc}). We thank the staff at BIFI and ICCAEx
supercomputing centers for their assistance.
\end{acknowledgments}

\appendix

\section{Multispin coding}

We have simulated the Metropolis dynamics through two different
multispin codings: MUlti SAmple multispin coding (MUSA) and MUlti SIte
multispin coding (MUSI) \cite{fernandez:15}.  

The MUSA algorithm is based on the representation of many sample
systems in a single computer word (128 bits in our implementation),
i.e. each bit represents a different sample; instead, the MUSI one
represents many spins of the same replica in a single computer word
(256 bits in our implementation). Indeed, the code implementing MUSA
is much simpler and thus it was adequate for the first stages of the
project. On the Intel(R) Xeon(R) E5-2680v3  processors of the
\emph{Cierzo} cluster, our MUSA code simulates 24 replicas of the same
sample at a rate of 12 picoseconds per spin flip (performance is
optimal with this configuration because the memory-consuming coupling
matrix is shared by the 24 replicas). Furthermore, the efficiency of
MUSA algorithm does not depend on the choice of boundary conditions,
either Open or Periodic.

On the other hand, the MUSI code has longer development times, but is
significantly faster than MUSA [the lower the temperature, the faster:
  MUSI codes update $\sim\exp(4/ T)$ spins with a single random
  number~\cite{fernandez:15}]. Indeed, at our highest temperature
$T=1.1$, on the E5-2680v3 processors, our MUSI code simulates 24
replicas at an overall rate of 8 picoseconds per spin
flip. Unfortunately, for open boundary conditions, spins on the top
(or bottom) layer have only 5 neighbors, which implies that one can
only update $\sim\exp(2/T)$ spins with a single random number. Hence,
we have implemented MUSI only for periodic boundary conditions.
\section{Statistical errors, samples and replicas}

We have computed $C_4\left(T,\vec r,t\right)$ [see Eq.~\eqref{fC4}]
at times $t=\text{integer-part-of }2^{i/4}$. For the
estimation of the integrals $I_k(T,t)$ [see Eq.~\eqref{sum_estimator}]
we have followed the methods explained in~\cite{fernandez:19}.

After a time $t^*$ the correlation length $\xi_{12}^{\parallel}$ does
not show any dependence of time, implying that thermal equilibrium has
been reached (see Fig.~\ref{fig:aging_rate}).  In the calculation of
Eq.~\eqref{fC4} at equilibrium there is no reason
to take the two real replicas at the same time $t$ and we can gain
statistics averaging over pairs of times $(t_1,t_2)$ both larger than
the safe equilibration threshold time $t^*$.

The choice of the optimal number of replicas $\NR$ and samples $\NS$
was chosen in order to minimize the final errors of the correlation
length $\xi_{12}^{\parallel}(t)$, given a fixed computer effort $E =
\NR\NS$.  Indeed, the variance (or squared error) in $\xi_{12}$
approximately follows this behavior in the off-equilibrium
regime~\cite{janus:18}:
\begin{equation}
\varDelta(\NS,\NR) = \left[ \sigmaS^2 + \sigmaR^2\left( \frac{2}{\NR(\NR-1)}\right)^{x}  \right]\frac{1}{\NS}\,,
\label{eq:error_C4}
\end{equation}
where the exponent $x$ takes a value in the range $0.5<x<1$,
$\sigmaS^2$ and $\sigmaR^2$ are (respectively) the sample and thermal
contributions to the variance and $\NR (\NR-1)/2$ is the number of
distinct pairs of replica indices for calculating $C_4(\vec r,t)$, see
Eq.~\eqref{fC4}. Clearly, we need to find a
compromise by minimizing the (squared) error achievable for a fixed
numerical effort $E=\NR \NS$, which results into an optimal value
\begin{equation}\label{eq:optimal-NR}
\NR^*\approx \left[2^x(2x-1)\frac{\sigmaR^2}{\sigmaS^2}\right]^{1/(2x)}\,,
\end{equation}
[the result is approximated because we simplified the algebra as
$\NR(\NR-1)\approx \NR^2$].
\begin{figure}[t]
\centering
\includegraphics[width=\columnwidth]{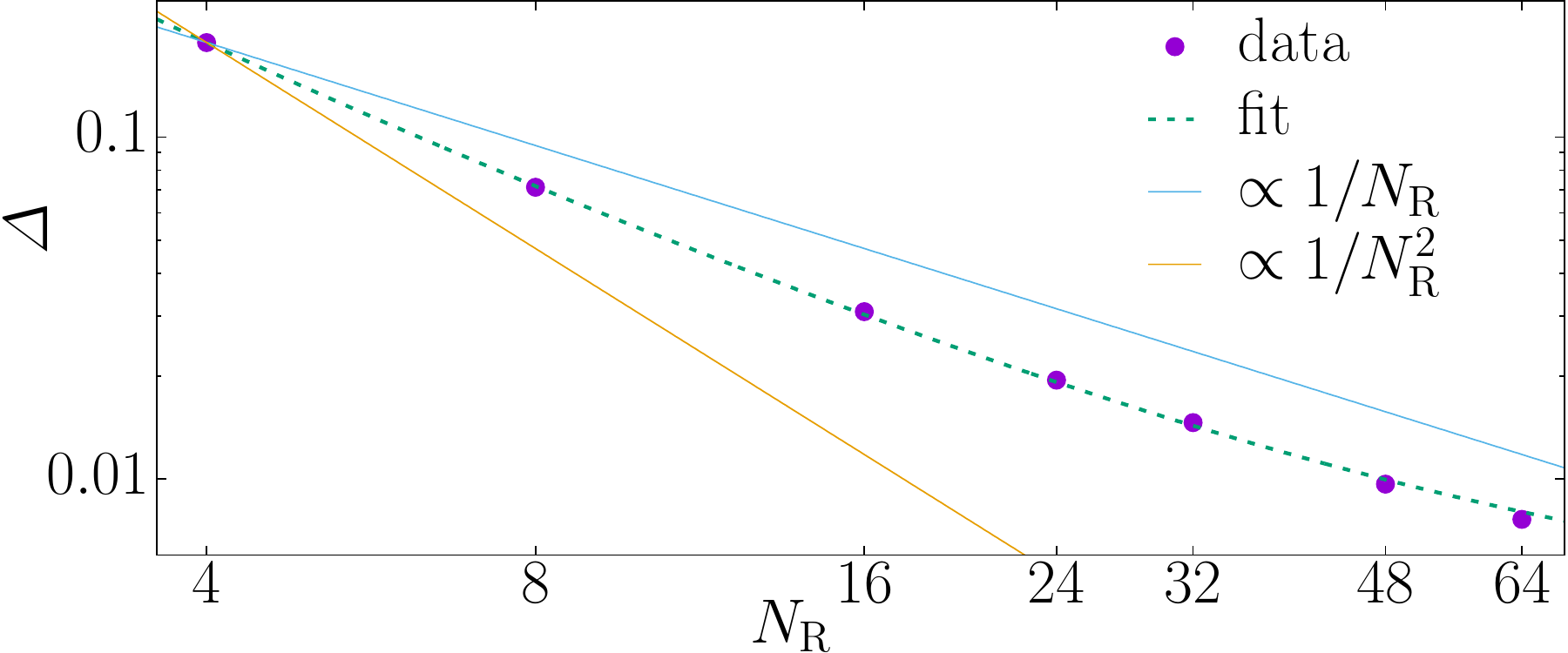}
\caption{\label{fig:optimal_NR} Squared statistical error for
  \mbox{$\xi_{12}(\Tc,t\!=\!2^{22},L_z\!=\!4)$}, as computed from a set of
  $\NS=128$ samples and $\NR$ replicas, versus $\NR$. The dashed line
  is a fit to Eq.~\eqref{eq:error_C4}. The relevant quantities
  extracted from the fit are $\sigmaR^2/\sigmaS^2\approx 156$ and
  $x\approx 0.65$. For reference, we show with continuous lines the
  two extremal behaviors, namely $x=1$ (with $\varDelta\propto
  1/\NR^2$) and $x=0.5$ (with $\varDelta\propto 1/\NR$). Because
  $\sigmaR^2\gg\sigmaS^2$, $\varDelta$ shows an intermediate behavior
  for small $\NR$. However, when $\NR>30$ the contribution of thermal
  fluctuations to the final error becomes comparable to the sample contribution
  and there is little gain in further increasing $\NR$.}
\end{figure}

At this point, we needed to estimate the ratio $\sigmaR^2/\sigmaS^2$,
as well as the exponent $x$. In order to do so, we carried out short
MUSI runs with $t=2^{22}$ at $\Tc$, for thickness $ L_z=4$, with
$\NR^{\text{tot}}=72$ and $\NS=128$. We randomly extracted
$\NR=4,8,16,24,32,48$ and $64$ replicas out of the ensemble of
$\NR^{\text{tot}}$ possibilities, and computed $\xi_{12}$ and its
squared error $\varDelta(\NS,\NR)$ with the jackknife, see
e.g. Ref.~\cite{amit:05} (we computed jackknife blocks over the samples).
In order to stabilize the estimation of $\varDelta$ we averaged over 20
random extractions of the $\NR$ replicas. The obtained
$\varDelta(\NR,\NS)$ are shown in Fig.~\ref{fig:optimal_NR} with our fit
to Eq.~\eqref{eq:error_C4}.

The resulting optimal value is $\NR^*\approx 29.4$ [the approximation
  in Eq.~\eqref{eq:optimal-NR} predicts 27.3]. However, by plugging
$\NS=E/\NR$ in Eq.~\eqref{eq:error_C4} and varying $\NR$ while
keeping $E$ fixed, we observed that the minimum at $\NR^*$ is quite
broad, which is fortunate because the value that optimizes the
performance of our MUSA code on the \emph{Cierzo} processors is
$\NR=24$.

Our final choices are as follows. With our (more flexible) MUSI code
we simulated $\NS=120$ independent samples, each with $\NR=32$
replicas.  In the MUSA case, we simulated 4 independent runs of 128
different samples and $\NR=24$ real replicas.
There is a caveat, though. The MUSA algorithm, sharing, by
construction, the random numbers for all the samples in a computer
word, could introduce some statistical correlation between different
samples. We initially checked the statistical correlation comparing
the error determination either assuming $\NS=512$ independent samples
or 4 independent blocks of 128 samples. Although with 4 sets the error
determination is very imprecise, we found no significant signal of
correlations. Furthermore, as soon as the MUSI algorithm was
implemented, we checked carefully the real statistical sample
independency by comparing the statistical errors for our observables
as computed with the two algorithms.  After this comparison, we found
consistent the computation of errors under the hypothesis that the 512
samples in the four MUSA simulations are statistically independent. In
fact, the independence hypothesis seems to systematically
underestimate errors only for $C_4(r,t)$ at distance $r=1$, and (probably)
for $r=2$ as well. The effect of this error underestimation can be
observed in global magnitudes such as $\xi_{12}^{\parallel}$ only for
very short times ($t<20$) when $\xi_{12}^{\parallel}$ itself is very
small. Hence, we have decided to accept the independence hypothesis in
our error computations for MUSA simulations.

\begin{figure}
\includegraphics[width=\columnwidth]{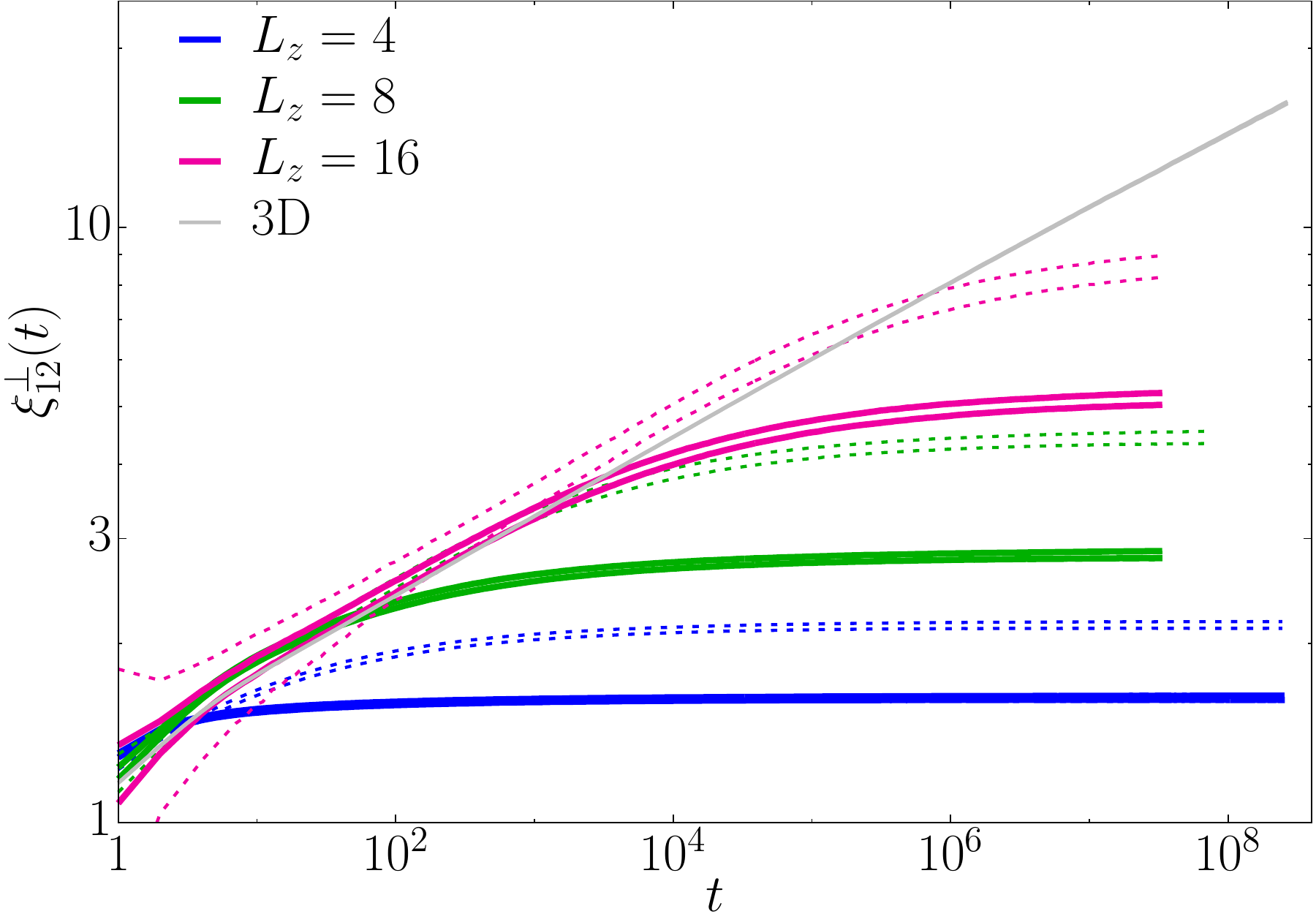}
\caption{\label{fig:FBC_transv} Growth of the transversal correlation
  length $\xi_{12}^{\perp}(T,t)$ at $T=0.98$ as function of the
  waiting time $t$ in scale log-log. The PBC (OBC) case is depicted in
  solid (dashed) lines.}
\end{figure}

\begin{figure}
\includegraphics[width=\columnwidth]{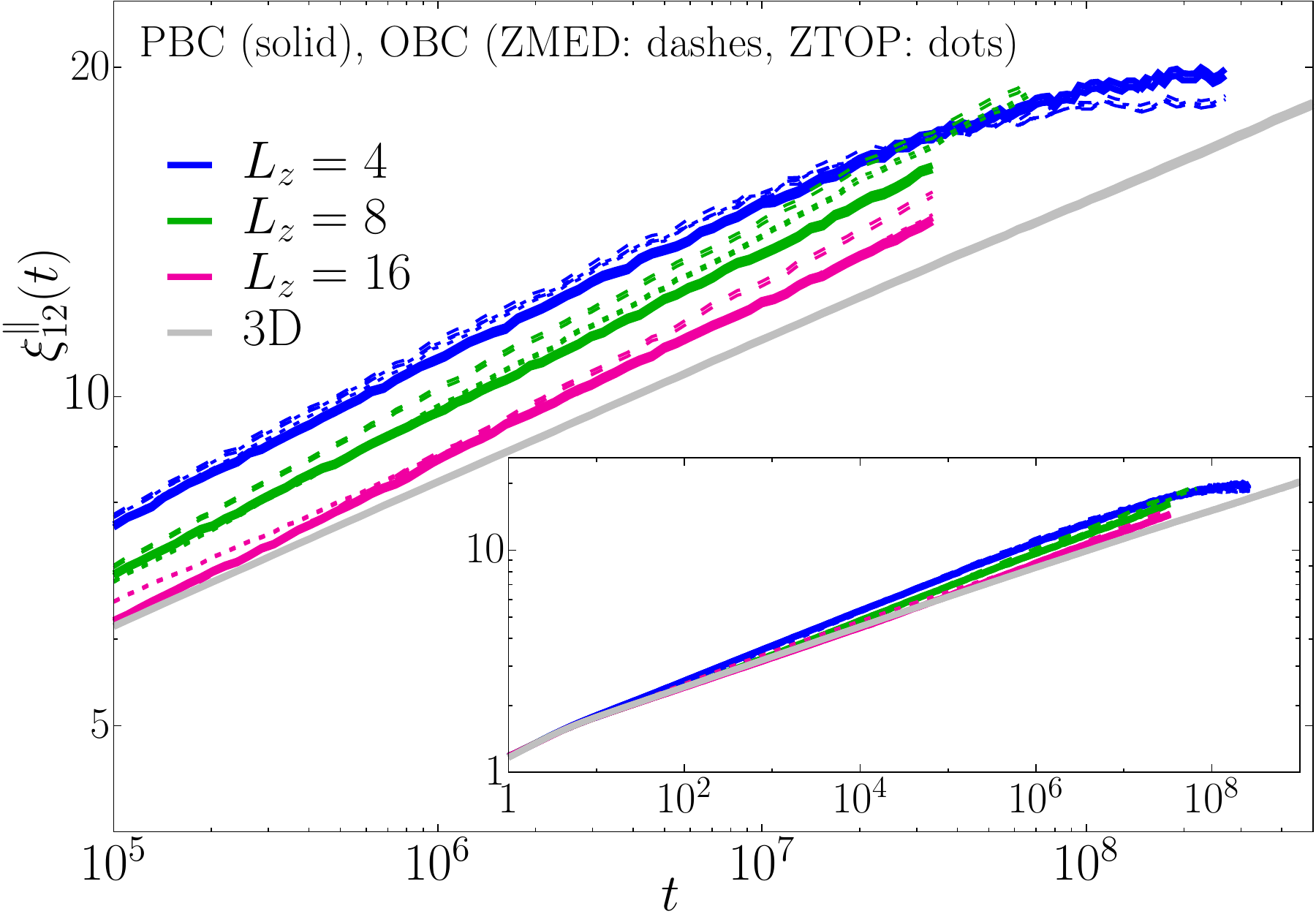}
\caption{\label{fig:boundary_condition} Growth of the longitudinal
  correlation length $\xi_{12}^{\parallel}(T,t)$ at $T=0.98$ as
  function of the waiting time $t$ as computed with periodic boundary
  conditions (PBC) or with open boundary conditions (OBC) on the
  central layer (ZMED) or on the top layer (ZTOP).  The \textbf{inset}
  shows the data from the main panel in the full time-range of our
  simulations.}
\end{figure}\begin{figure}

\includegraphics[width=\columnwidth]{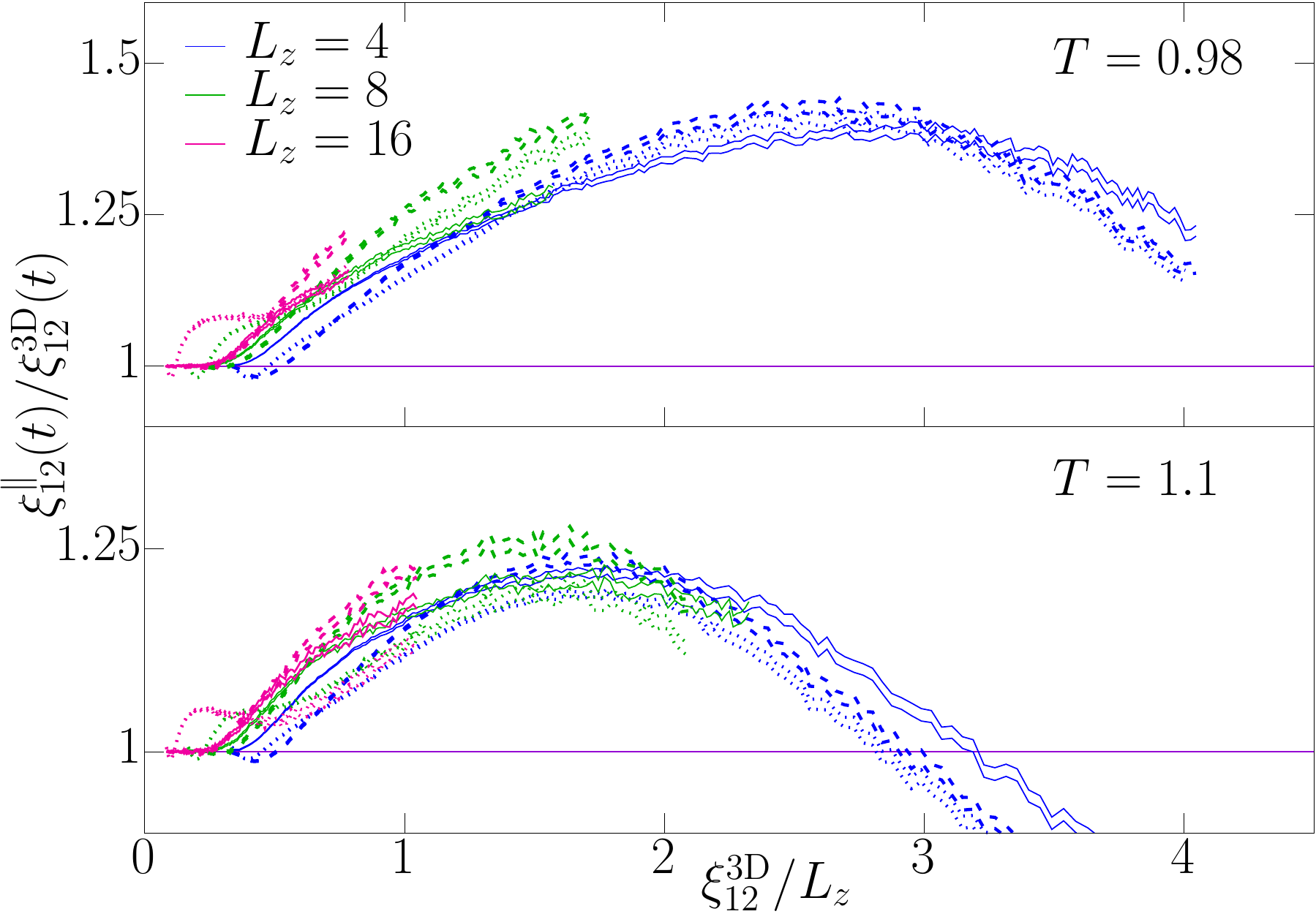}
\caption{\label{fig:FBC_scaling} Dynamical scale-invariance for the
  dimensionless quantity $\xi_{12}^\parallel(t)/\xi_{12}^\mathrm{3D}(t)$ as a function of the
  rescaled bulk length $\xi_{12}^\mathrm{3D}(t)/L_z$, for the PBC case
  (solid), for the OBC central-layer (dashes) and for the OBC
  external-layer (dots) at temperatures $T=0.98$ ({\bf top}) and
$T=1.1$ ({\bf bottom}).}
\end{figure}

\section{Boundary conditions}
In the phenomenology of glassy films, the transversal saturation of
$\xi_{12}^{\perp}$ activates the dimensional crossover and so, the
boundary conditions could play a physical relevant role. In order to
assess the effect of the boundary conditions, we carried out MUSA
simulations with both Open (OBC) and Periodic Boundary Conditions
(PBC) for several temperatures and $L_z$'s (recall that $L_z$ is the
film thickness).  Exploiting the same kind of analysis introduced in
the main text, we found that our main results are not dependent on
boundary conditions.

Regarding the sum estimator $\xi_{12}^\perp$ defined in
Eq.~\eqref{sum_estimator}, for OBC and computing the correlations from the
bottom layer at $z\!=\!0$, we can extend the sum up to $L_z-1$. Hence, by
construction, $\xi_{12}^{\perp}$ is larger for OBC than for PBC (see
Fig. ~\ref{fig:FBC_transv}).

As for the comparison of the parallel dynamics, in the case of OBC we
need to face the possibility of layer dependence. However,
Fig.~\ref{fig:boundary_condition} tells us that the differences
between $\xi_{12}^{\parallel}(T,t)$ as computed for the top layer and
the central layer are tiny (and the difference with the PBC result is
tiny as well), although our data are accurate enough to resolve the
difference. In fact, see Fig.~\ref{fig:FBC_scaling}, the
layer-dependence with OBC makes slightly more complicated the analysis
of scaling functions.

\section{Renormalization Group}

We decomposed our system in boxes of size of $ L_z^3$ and we rescaled
the overlap field as
\begin{equation}
Q^{(a,b)}(\vec X,t) = \frac{1}{L_z^3}\sum_{r_1,r_2,r_3=0}^{L_z-1} q^{(a,b)}(\vec r+L_z\vec X,t)\,,
\end{equation}
\newpage
\noindent and we defined the correlation function in the (2D) renormalized lattice as:
\smallskip
\begin{equation}
C_4^\text{RG}(T,\vec R,t)=\overline{\langle Q^{(a,b)}(\vec X, t) Q^{(a,b)}(\vec X+\vec R, t)\rangle_T}\,.
\end{equation}
We gain statistics by averaging over all the $L_z^3$ possible starting
position of the boxes and all pairs of different replicas. The
estimate of the correlation length was done as well through the
integral estimators defined in the main text. Specifically, we
computed the integrals
\begin{equation}
I^\text{RG}_k(T,t) = \int_0^{\infty} \mathrm{d}R \, R^k\, C_4^\text{RG}(T,R,t)\,,
\end{equation}
and we estimated the correlation length as
\begin{equation}
\xi_{k,k+1}^\text{RG}(T,t)=
I^\text{RG}_{k+1}(T,t)/I^\text{RG}_k(T,t)\,.
\end{equation}

\bibliographystyle{apsrev4-1}
%

\end{document}